\def \etal {{\it et al. }}
\def \rd {{\rm d}}
\def\cP{{\cal P}}

\def \kpc {h^{-1}{\rm kpc}}

\documentstyle[aas2pp4]{article}

\slugcomment{CfPA 98-th-03}

\begin{document}

\title{Deprojection of Rich Cluster Images}

\author{S. Zaroubi\altaffilmark{1,2}, G. Squires\altaffilmark{3}, 
Y. Hoffman\altaffilmark{1,4} and J. Silk\altaffilmark{3,5}}
\altaffiltext{1}{Racah Institute of Physics, The Hebrew University, 
Jerusalem 91904, Israel}
\altaffiltext{2}{Department of Medical Biophysics and Nuclear Medicine, 
Hadassah Hebrew University Hospital, P.O. Box 12000, Jerusalem, 91120, Israel}
\altaffiltext{3}{Center for Particle Astrophysics, University of
California, Berkeley, CA 94720, U.S.A.}
\altaffiltext{4}{Kapteyn Astronomical Institute, University of Groningen, P.O. 
Box 800,  9700 AV Groningen, The Netherlands}
\altaffiltext{5}{Departments of Astronomy and Physics, University of
California, Berkeley, CA 94720, U.S.A.}

\begin{abstract}
  We consider a general method of deprojecting two-dimensional images to
reconstruct the three dimensional structure of the projected object,
assuming axial symmetry.  The method consists of the application of the
Fourier Slice Theorem to the general case where the axis of symmetry is
not necessarily perpendicular to the line of sight, and is based on an
extrapolation of the image Fourier transform into the so-called cone of
ignorance. The method is specifically designed for the deprojection of
X-ray, Sunyaev-Zeldovich (SZ) and gravitational lensing maps of rich
clusters of galaxies. For known values of the Hubble constant $H_0$ and
inclination angle, the quality of the projection depends on how exact is
the extrapolation in the cone of ignorance. In the case where the axis
of symmetry is perpendicular to the line of sight and the image is
noise-free, the deprojection is exact.  Given an assumed value of $H_0$,
the inclination angle can be found by matching the deprojected structure
out of two different images of a given cluster, e.g., SZ and X-ray
maps. However, this solution is degenerate with respect to its
dependence on the assumed $H_0$, and a third independent image of the
given cluster is needed to determine $H_0$ as well. The application of
the deprojection algorithm to upcoming SZ, X-ray and weak lensing
projected mass images of clusters will serve to determine the structure
of rich clusters, the value of $H_0$, and place constraints on the physics
of the intra-cluster gas and its relation to the total mass
distribution.  The method is demonstrated using a simple analytic model
for cluster dark matter and gas distributions, and is
shown to provide a stable and unique reconstruction of the cluster 3D
structure.
\end{abstract}


\section{Introduction}

Uncovering the three-dimensional (3D) structure {}from observations of
projected quantities is a fundamental problem of astrophysics
research. The 3D structure of rich galaxy clusters is particularly
interesting, as the cluster geometry impacts on determinations of the
Hubble constant {}from X-ray and Sunyaev-Zeldovich (SZ) measurements,
cluster mass and baryon fraction determinations, and the underlying
galaxy orbit structure.

The most straightforward approach when dealing with projected quantities
is to assume a spherical symmetry.  The assumption is often applied
within the framework of parametric models, for example the standard
$\beta$-model (e.g., \cite{cavaliere76}; see \cite{rephaeli95} for a
review), or a non-parametric geometric approach (\cite{fabian81}). Such
simplifications can lead to biases in whatever quantity is being extracted
{}from the data, especially when the systems under scrutiny might be
biased to a certain geometry such as, for example, being detected since
they are preferably elongated along the line of sight (hereafter
LOS). This bias can be considerable as was illustrated in the particular
case of merging clusters (\cite{roet97}).

Here, we restrict ourselves to the problem of reconstructing the 3D
structure {}from maps of the X-ray emission or the SZ decrement (say),
assuming the underlying 3D distribution is axially symmetric. We show
that the Fourier Slice Theorem (\cite{rybicki87}), can be extended to
general case of an arbitrary angle and the deprojection can be done by
applying a simple approximation. In the case of astronomical images the
inclination angle is  unknown.  However, with the availability of
SZ, X-ray and lensing maps, we can uniquely determine the cluster
inclination angle by comparing the deprojections of the various
images. This result enables us to ascertain the $H_0$ dependence in the
more general case where an axial symmetry is assumed.  We apply this
technique for the reconstruction of the 3D cluster structure to
simulated SZ and X-ray images in an analytical cluster model.

\section{Deprojection} \label{sec:deprojmethod}

The deprojection of axially symmetric quantities is a classical problem
(e.g., \cite{lucy74}), considered for determining the 3D stellar orbit
structure in elliptical galaxies (e.g., \cite{dehnen93};
\cite{dehnen94}; \cite{bdi90}), and in determining cluster 3D structure
{}from X-ray observations (\cite{fabricant84}).

We deproject cluster images as follows. We adopt the convention that bold-face
symbols denote 3D quantities (e.g., ${\bf k} = (k_x, k_y, k_z)$).  Let
the observer's coordinate system be defined with the Cartesian axes
$(x^\prime,y^\prime,z^\prime)$, with the $z^\prime$ axis aligned with
the LOS.  Denote the (cluster) source function coordinate
system by the axes $(x,y,z)$ where the $z$-axis is the cluster symmetry
axis, forming an angle $\theta_i$ with respect to the LOS.  Let
$I(x^\prime,y^\prime)=\int\lambda^\prime({\bf x^\prime})\,dz^\prime$
denote a projected quantity (image) of the source function $\lambda$.
The 3D Fourier transform (FT) of the source function is related to the
image transform by
\begin{eqnarray}
\lambda_{\bf k^\prime}^\prime(k_x^\prime,k_y^\prime,0)   
  &  = &\int e^{[-i(k^\prime_x x^\prime + k^\prime_y y^\prime)]}
I(x^\prime,y^\prime) \, \rd x^\prime \, \rd y^\prime  \nonumber \\
  & =  & I_{k^\prime}(k_x^\prime,k_y^\prime).  \label{eqn:simagereln} 
\end{eqnarray}

The relations are easily seen in the case $\theta_i = 90^\circ$:
since the cluster has a rotational symmetry, in the cluster frame
$\lambda({\bf r}) = \lambda(r,z)$ and its FT is $\lambda_{\bf k}({\bf
k})=\lambda_{\bf k}(k,k_z)$ where $k=\sqrt{k^2_x + k^2_y}$.  Now 
$k=\sqrt{k^{\prime 2}_z + k^{\prime 2}_y}$ and $k_z=
k^\prime_x$, and the relation between the FT of the source function in the
observer and cluster rest frames is $\lambda^\prime_{\bf k^\prime}({\bf k'}) = 
        \lambda_{\bf k}\biggl( \sqrt{ k^{\prime 2}_y + k^{\prime 2}_z },
        k^\prime_x \biggr)$.
Deprojection is obtained by inverse transforming,
and using equation (\ref{eqn:simagereln}), 
to obtain
\begin{equation}
\lambda(r,z) = {\frac{1}{(2 \pi)^2}} \int e^{i k_z z}  I_{k^\prime}
        (k, k_z) J_0(k\, r) 
k\,dk\,dk_z. 
\label{eqn:fouriersliceeqn}
\end{equation}

This particular case where the symmetry axis is perpendicular to the
LOS is an exact application of the
Fourier Slice Theorem.  The general case is easy to obtain. For
arbitrary inclination angles, equation (\ref{eqn:simagereln}) holds. The
expression relating the FT of the source function in the observer and
cluster rest frames is obtained simple by coordinate rotation, so that
\begin{eqnarray}
                   \lambda_{k^\prime}^\prime({\bf k^\prime})&  = &
                   \lambda_k  \biggl( \sqrt{(-k_z^\prime \sin \theta_i +
k_x^\prime
                   \cos \theta_ i)^2
                  + k_y^{\prime 2}},  \nonumber \\
& &     \;\;\;\; \;\;\;\; k_z^\prime \cos \theta_i + k_x^\prime \sin
                   \theta_i\biggr) \nonumber \\
                 & = &\lambda_{k}(k,k_z). \label{eqn:lambdareln}
\end{eqnarray}
where the last equality again holds due to axial symmetry. Inverse
transforming, we find the desired expression for the source
function in real space
\newpage

\begin{eqnarray}
\lambda^{\rm deproj}(r,z) = {\frac{1}{(2 \pi)^2}} \int  
	k\,dk\,dk_z \,e^{i k_z z}  J_0(k\, r)  \times \nonumber \\ 
I_{k^\prime}\biggl( \frac{k_z}{\sin \theta_i}, 
	\sqrt{k^2-k^2_z\cot^2 \theta_i}\biggr) . \label{eqn:maineqn}
\end{eqnarray}

For wave-vectors $ k < |k_z| \cot \theta_i$ the argument of the image FT
becomes imaginary.  This defines what has been dubbed the ``cone of
ignorance'' (\cite{rybicki87}; \cite{gerhard96}) inside which there is
no information on the 3D structure in the image.  Generally, $
\lambda^{\rm deproj}(r,z) \neq \lambda(r,z)$, and the difference between
the two depends on the power hidden in this region. For clusters aligned
along the LOS, this region completely covers k-space, and the method
fails. In the other extreme, $\theta_i = 90^\circ$, all of the
information required for a unique deprojection is contained within the
image.

The implementation of this method is straightforward, assuming for the
moment that the inclination angle is known. The image FT is
evaluated at wavevectors  
\begin{equation}
k_x^\prime  =  k_z/\sin \theta_i  \;\;\; {\rm and} \;\;\; k_y^\prime  =
\sqrt{k^2 - k_z^2 \cot \theta_i^2} \label{eqn:wavevectors}
\end{equation}
and the inverse transform applied. Numerically, this requires some
interpolation between the grid points of a finite FT.
For moderately smooth underlying cluster distributions, this
should not pose difficulties although this can be rigorously tested with
numerical cluster simulations, for example. We consider a simple
application in \S \ref{sec:toymodel}.

The cone of ignorance is a problem for the uniqueness of the
deprojection.  Palmer (1994) shows that any 3D axisymmetric source
function can be uniquely recovered via deprojection if it can be
expanded as a finite sum of spherical harmonics. However, Gerhard \&
Binney (1996) have constructed an example that obeys this theorem and
lurk only within the cone of ignorance, while still having a physically
interesting density distribution (a disk like structure).  This type of
degeneracy has limited the utility of this technique in recovering the
3D density distribution in galaxies, where this type of structure is
plausible. However, Romanowsky \& Kochanek (1997) employed a numerical
regularization method deal with information hidden in the cone of
ignorance, and constrained physically acceptable models by comparing
with observed stellar velocity dispersion.

Clusters are more favorable systems for this method.  Often cluster
images are elongated along an (approximate) axis of symmetry.  In a
bottom-up clustering scenario, cluster formation proceeds by an ongoing
merging process, and
one expects clusters not to be highly flattened, and to
be preferentially elongated objects.  Moreover, clusters form via
dissipationless clustering and therefore lack sharp discontinuities and
irregularities, resulting in a relatively smooth and regular Fourier
space behavior. Given this, at least some of the power resides outside
the cone of ignorance, which permits extrapolation into this
region. Moreover observations of the weak lensing effect, the X-ray
surface brightness and the SZ decrement provide projections of different
powers of the gas and mass density distributions so that only very
contrived models will be degenerate in all three images. The idea is
then that by combining X-ray, SZ and lensing observations, we can hope
to {\em determine} the inclination angle by comparing the deprojection
of the various images -- a variant of this idea was considered some time
ago in an analysis of the X-ray emission and the galaxy distribution in
the cluster A2256 (\cite{fabricant84}).

\section{Noise properties} \label{sec:noisecorr}

To compare the deprojection {}from various wavelengths statistically, we
need to characterize the noise properties of the source functions.  As
with all deprojection methods, the noise is correlated in the resulting
3D map.  Suppose the image is pure white noise, 
with rms $\sigma$, and hence
$
\langle I (x_1^\prime, y_1^\prime) 
I (x_2^\prime, y_2^\prime) \rangle = 
        (2 \pi)^2 \delta(x_1^\prime - x_2^\prime)
\delta(y_1^\prime - y_2^\prime) \; \sigma^2$. 
The noise correlation function in the deprojected image is given by
\begin{eqnarray}
\langle \lambda(r_1, z_1) \lambda (r_2,z_2) \rangle = \sigma^2 \,
        \frac{ \sin\theta_i }{ (2 \pi)^2} 
        \int  k \, dk_z \, dk \, e^{ i k_z (z_1 - z_2)} \times \nonumber \\
	 \sqrt{ k^2 - k_z^2 \cot\theta_i} \,
        J_0(k r_1) \, J_0(k r_2).
\end{eqnarray}
Not surprisingly, the high frequency components dominate and hence some
smoothing must be employed.  The analytic nature of the correlation
matrix is an attractive feature of the method. This allows a direct
statistical comparison of the source functions determined independently
{}from SZ and X-ray observations, and alleviates the need for
simulations to determine the statistical significance of features found
in the deprojections.

A further nice feature of the method is that any isotropic smoothing
function applied to the image will depend only on the modulus of the
radial wavevector, $|{\bf k}|$, in the deprojection.  This can easily be
seen since the identification of coordinates between the observer's
frame and the frame of reference of the source function given in
equation (\ref{eqn:wavevectors}) yields $k^{\prime 2} \rightarrow
k^2+k_z^2 = | {\bf k}^2 |$.  A simple and practical Gaussian smoothing
filter gives a particularly simple correlation of the noise $\langle
\lambda(r)^2 \rangle \propto r^{-1}$.

\section{Practical Application} \label{sec:toymodel}

To illustrate the method, we consider a simple analytical model for the
cluster gas distribution. We use a simple gas density profile, $
\rho_g(x,y,z) \propto [ r_c^2 + (x^2 + y^2)/a^2 + z^2/c^2]^{-1}$ and
create simulations of SZ and X-ray observations for a prototype
cluster with the following characteristics: $r_c = 100 \kpc$, $T =
7$~keV, central density of $\rho_{g0} = 0.01 {\rm \; cm}^{-3}$, axis
ratio of 3:1 and an inclination angle of $\theta_i =60^\circ$.  White
noise is added to mimic levels of ROSAT PSPC observations, and SZ maps
of \cite{cjg97}.

The procedure then to determine the cluster 3D structure is as follows:
given X-ray and SZ images, we
apply a smoothing filter to reduce the noise
levels in the images. Motivated by the high degree of noise removal
necessary for this type of application, one of us (Hoffman 1998) has
proposed a method for noise removal by wavelet thresholding. This method
has the advantage of superior noise suppression over simple linear
filtering, while maintaining spatial resolution. After denoising, an
inclination angle is assumed, and the FT of the images formed. To
perform the deprojection, we
employ equation (\ref{eqn:maineqn}), and
evaluate the FT at wavevectors given by equation
(\ref{eqn:wavevectors}). This requires some interpolation {}from the
finite image FT, and we employed cubic splines.

For smooth cluster density profiles, the image FT is quite regular and
the extrapolation into the cone of ignorance can be done robustly. For
this example, we employ a relatively simple linear extrapolation, with
the slope being taken {}from the tangent to the contours at the cone
boundary. If significant substructure exists, the worry is that
this sort of extrapolation would smooth over the structure.  Clearly,
more sophisticated approaches could be adopted as the data warrants,
however this simple extrapolation seems to work well for relatively
smooth cluster profiles. We are currently undertaking a more thorough
investigation using cluster SPH simulations, and we expect that the
simplistic approach adopted here will not be the final word on the
subject.

Having  evaluated the image FT at the wavevectors in equation
(\ref{eqn:wavevectors}) and extrapolated into the cone of ignorance, we
are in position to perform the deprojection.  If the inclination angle
is known, the reconstructed 3D structure of the source functions
$\lambda_{SZ}(r,z)$ and $\lambda_X(r,z)$ is excellent. 
Of course the fact we have ignored heretofore is that the method
requires a priori that the inclination angle is known, whereas this is
not the case for astronomical observations of clusters. For any one
image, all one can do is assume some angle. 

\begin{figure}
\plotone{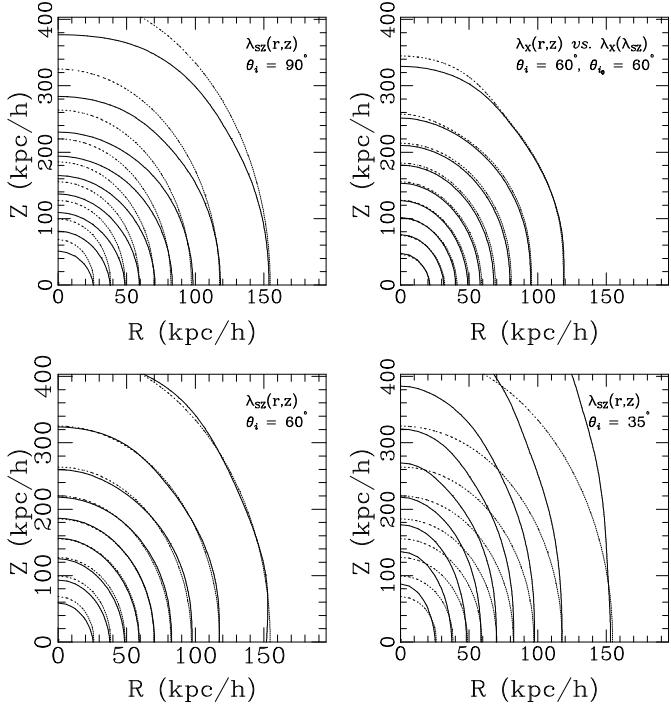}
\caption{Reconstruction of $\lambda(r)$ at fixed $z$ for different
inclinations, showing actual and reconstructed images. The top left, and
bottom left and right panels show the SZ deprojection (solid lines),
compared with the true underlying source function (dashed line), assuming three
different inclination angles. The top right panel shows the gas density
inferred {}from the X-ray and SZ maps independently, with the correct value
of the inclination angle used in both cases. The agreement in the X-ray
and SZ deprojections is quite good, if the inclination angle is
known. This suggests this is a viable method for determining the cluster
geometry by requiring consistency between the two deprojections.}
\label{fig:unknownangle}
\end{figure}

The error made assuming the incorrect inclination angle is displayed in
Figure \ref{fig:unknownangle}. The top left and the bottom left and
right panels show the deprojection of the SZ image compared with the
actual SZ source function for three assumed inclination angles.  In the
case where $ \theta{^{\rm assumed}_i} \neq \theta_i$, a substantial
discrepancy between the actual and deprojected source function is found.
This is to be compared with the good agreement found when the correct
$\theta_i$ is assumed. To further investigate the robustness of the
method the following comparison was made (Figure \ref{fig:unknownangle},
top-right panel): using the correct inclination angle, the deprojected
SZ source function had been used to to calculate the X-ray source
function, which was then compared with the actual one. Again,  very
good agreement is found.

This comparison of the various observables is admittedly simplistic and the
procedure for real observations will necessarily be more involved. For
example, the current best resolution and highest signal-to-noise SZ
measurements being obtained at BIMA and OVRO are interferometric
(Carlstrom, Joy \& Grego 1997). Since the SZ data is in
Fourier space, with irregular spacing, the algorithm as presented here
of deprojecting two or more images does not strictly apply. However, it
is easy to compare the X-ray and SZ data in some other fashion by, for
example  deprojecting the X-ray image, constructing the
$\lambda_{SZ}(\lambda_X)$ source function {}from the X-ray deprojection,
project this source function into the SZ observational plane, and there
make the comparison. This type of application will need to be tailored
to the particular data set in hand and tested by numerical cluster
simulations.

Note that weak lensing observations provide a further probe into the
cluster 3D structure. However, to compare the observable {}from weak
lensing, the total mass surface density, with X-ray and SZ observations,
which probe the gas density and temperature distributions, we
need to appeal to an assumption for the cluster dynamical state.
Assuming the gas is in hydrostatic equilibrium in the cluster potential
well, the gas pressure, $P$, is related to the potential, $\phi$, by
$\nabla P = - \rho_g \nabla \phi(r)$ and the total mass density is
determined via the Poisson equation $\nabla^2 \phi = 4 \pi G \rho$. This
illustrates the procedure for comparing the maps: given SZ and X-ray
images, one deprojects to determine the gas density. {}From weak lensing
observations, one determines the total mass density. Comparing the
deprojections by assuming hydrostatic equilibrium will then yield
further insight into the dynamical state of the cluster, and the
relative distributions of the dark matter and the gas.

\section{Cosmological Parameter Dependence} \label{sec:hubble}

The deprojection of images of objects at cosmological
distances is further complicated by the $H_0$ dependence, but it also provides
an interesting and practical way for determining the value of $H_0$.
We first recall the Hubble constant dependence on the X-ray and SZ maps.
Let the observer's coordinate systems be denoted by $(\theta_x,\theta_y,
z)$, where $z$ is the cluster redshift. Physical coordinates are given
by $\bar x' = D_A(z)\, \theta_x$, $\bar y' = D_A(z) \, \theta_y$ where
$D_A(z)$ is the angular diameter distance to the cluster. The Hubble
constant dependence can be factored out as $D_A(z) \propto h^{-1}$ and
hence we rescale the variables by $ \bar x' = h^{-1} x'$ (similarly for
y). Expressing the image and the object structure in these coordinates,
the Hubble constant dependence is explicit: the 2D FT of
the image yields an $h^{-2}$ scaling and the 3D inverse FT
gives an $h^3$ scaling.  It follows that the expression for
$\lambda(r,z)$ (equation \ref{eqn:lambdareln}) varies as $h$.

We can see this explicitly, adopting a model for the intracluster
gas. Let $\cP$ to be the projection operator and $\cP^{-1}$ be the
inverse. The X-ray source function (which is proportional to $\rho_g^2
\, T^{1/2}$) is given by $\lambda_x(r,z) = h \,
\cP^{-1}[I_x(\theta_x,\theta_y)]$ (and similarly for the SZ source
function, $\lambda_{sz} \propto \rho_g \, T$).  Then given a model that
relates $\lambda_x$ and $\lambda_{sz}$, one can solve for $h$. For
example, for a polytropic gas ($p\propto \rho_g^\gamma$ and
$T\propto \rho_g^\beta, \beta=\gamma-1$), one finds
\begin{equation}
h = \frac{ \left( \cP^{-1}[I_{sz}]\right)^{ \frac{4+\beta}{2-\beta} }
      }
      { \left( \cP^{-1}[I_x   ]\right)^{ \frac {2+\beta}{2-\beta}}  }
\end{equation}

\begin{figure}
\plotone{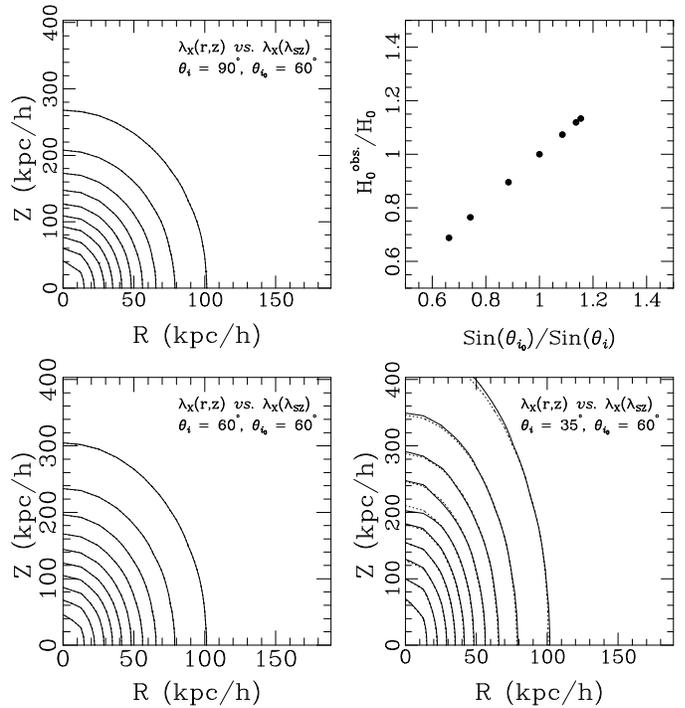}
\caption{The effect of the (unknown) inclination angle on the
  determinations of $H_0$. The bottom and top left panels compare the
  X-ray source function $(\propto \rho^2 \, T^{1/2})$ in dashed
  contours, with the prediction {}from the deprojection of the SZ map,
  using 3 different assumed inclination angles. Good agreement is
  obtained in the {\em shape} of the X-ray isophotes for any assumed
  inclination angle.  However, the amplitude is mismatched and this
  translates in an error on the inferred value of $H_0$, shown in the
  top right panel. The inferred value of $H_0$ scales with the
  inclination angle as $\sin\theta_i$.}
\label{fig:h0dep}
\end{figure}

 In the model employed here, the deprojection operator
scales with $\sin\theta_i$. Thus $\cP{^{-1}_i}\propto \sin\theta_i$, and
therefore {\em each} of the source functions {}from the X-ray and SZ
deprojection has a scaling with $h \sin\theta_i$. This is illustrated
in Figure \ref{fig:h0dep}. We created simulated noise-free SZ and X-ray
maps, with the density model described in \S \ref{sec:toymodel} and a
polytrope of $\gamma = 5/3$. We applied the deprojection to both images.
We took the deprojected $\lambda_{SZ}$, predicted the X-ray source
function, $\lambda_x$, and compared with the deprojection of the X-ray
image.  In the three panels of Figure \ref{fig:h0dep} we display the
results for three assumed inclination angles. In all cases,
the {\em shape} agreement is almost perfect, but the normalization
disagrees.

The relation between the true and assumed inclination angle and the true
and inferred Hubble constant is shown in the top right panel of Figure
\ref{fig:h0dep}. The basic result is that the value for $H_0$ inferred
scales with $\sin\theta_i$.  We have not succeeded in proving
this $h \sin\theta_i$ degeneracy in the general case of an
arbitrary 3D structure, but we speculate that this behavior
characterizes (at least to first order) cluster images in general. 

The $h\sin\theta_i$ degeneracy can be interpreted either as a positive
or negative feature of the method. The encouraging fact is that, at
least in the context of the model of the cluster gas used here,
requiring agreement between the deprojected source functions will place
joint constraints on $H_0$ and the inclination angle. The disappointing
aspect is that {}from these two maps alone, one can not constrain the
cluster geometry and $H_0$ independently.

\section {Conclusion} \label{sec:discussion}

We present a new algorithm for constructing the 3D structure of
axisymmetric objects {}from projected images.  The method is shown to be
especially suited to the deprojection of images of clusters of galaxies,
and is deonstrated to be stable in the presence of noise and quite a `wide'
cone of ignorance.  It is shown that given an assumed value of $H_0$ and
a model relating the gas density and temperature, the cluster
inclination angle can be solved given its SZ and X-ray maps. With
another image of the cluster, $H_0$ and the inclination
angle can be determined independently.  The application of
the method to realistic cluster simulations, and to SZ, X-ray and
weak lensing observations, will be presented in forthcoming
papers.

\newpage

\acknowledgements{We thank Ofer Lahav for insightful discussions.  This
research has been supported by a US-Israel BSF grant 94-185 (YH and JS),
by the Hebrew University S.A. Schonbrhnn Research Endowment Fund
(YH), and by the NSF (JS).}

{}

\end{document}